\documentstyle{elsart}

\def\tube{{\bf s}}
\def\occup{{\bf n}}
\def\tni{\theta(n_i)}
\def\conf{\occup;\tube}

\begin{document}
\input epsf
\begin{frontmatter}

\title{Repton model of gel electrophoresis in the long chain limit}

\author{M. Widom and I. Al-Lehyani}
\address{Department of Physics,
	Carnegie Mellon University,
        Pittsburgh, Pa. 15213}

\begin{abstract}
Reptation governs motion of long polymers through a confining
environment. Slack enters at the ends and diffuses along the polymer
as stored length. The rate at which stored length diffuses limits the
speed at which the chain can drift. This paper relates the rate of
stored length diffusion to the conformation of the tube within which
the polymer is confined. In the scaling limit of long polymer chains
and weak applied electric fields, holding the product of polymer
length times field finite, the tube length and stored length density
take on their zero-field values. The drift velocity then depends only
on the the polymer's end-to-end separation in the direction of the
field.
\end{abstract}
\end{frontmatter}

\section{Introduction}
\begin{figure}
\epsfysize=200pt \epsfbox{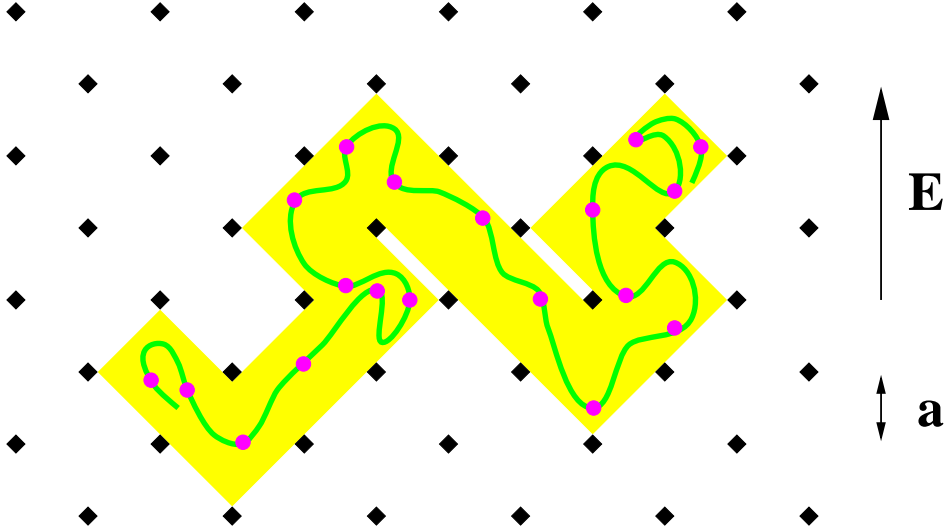}
\caption{
\label{gel.eps}
Polymer (solid curve) entangled in gel (squares) confined to tube
(shaded area). Dots spaced at intervals of $l_p$ are called reptons.}
\end{figure}

Gel electrophoresis separates charged polymers, such as sections of
DNA, according to their length.  Applied electric fields exert a
uniform force per unit length parallel to the field.  The gel contains
pores and fibers which tend to entrap the long polymers within
tubes~\cite{Edwards} (see figure~\ref{gel.eps}), impeding their drift.
Significant motion of the polymer occurs by the transport of stored
length through the tube, a process known as reptation~\cite{deGennes}.
As a result of entanglement of the polymer in the gel, the polymer
develops a complicated dependence of drift velocity on electric field
${\cal E}$ and total polymer length ${\cal L}=l_p N$. We denote the
polymer persistence length (typically of order 150-300 base pairs for
DNA) as $l_p$, and call $N$ the ``chain length'' of the polymer.

In the limit of weak field, the velocity is proportional to the
electric force through the Nernst-Einstein relation
\begin{equation}
\label{Nernst}
v_d = {{D}\over{k_B T}}(q N {\cal E}) .
\end{equation}
Here $(q N {\cal E})$ is the electric force on the polymer, with $q$ the
charge within a persistence length. The diffusion constant $D$ depends
on the chain length according to the prediction of de Gennes~\cite{deGennes}
\begin{equation}
\label{diff_constant}
D \sim N^{-2}.
\end{equation}
Comparing equations~(\ref{Nernst}) and~(\ref{diff_constant}), the weak
field drift velocity varies inversely with the chain length.
Short chains travel more quickly than long chains, and
chains may be sorted according to length based on their travel time
across the gel, or their travel distance within a given time.

Loss of length resolution, a serious impediment to electrophoresis as
a separation tool, occurs for any polymer above a certain length. For
DNA typical limits~\cite{pulsed} are of order 50 Kbp, with chain
length $N$ of order 150-300. Increasing these limits requires reducing
the electric field, thus increasing the run time, or using more
delicate gels.  Pulsed-field~\cite{pulsed} and other techniques push
the threshold length out yet further, reaching DNA lengths of order
$10^7$ bp, with chain length $N$ of order $10^5$.  From a mathematical
point of view, loss of length resolution implies a breakdown of the
Nernst-Einstein relation between drift velocity and zero-field
diffusion constant.  Understanding this problem requires further
investigation of the functional form of the drift velocity
$v_d(N,{\cal E})$.

\begin{figure}
\epsfysize=135pt \epsfbox{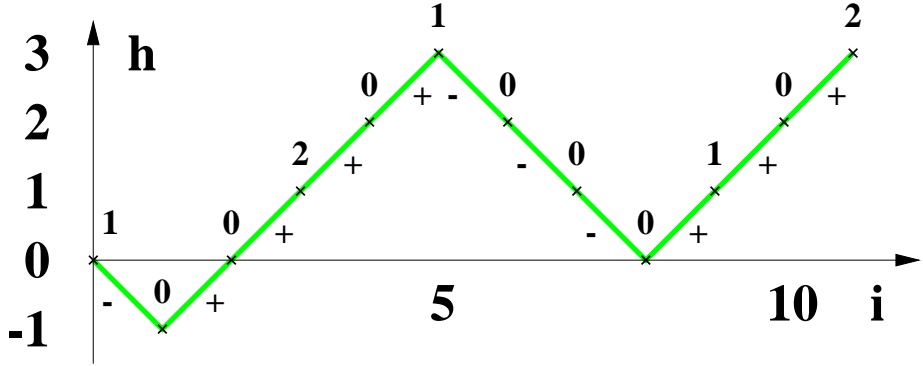}
\caption{
\label{repton.eps}
Repton model in the $(\conf)$ representation.  The numbers along the
chain denote $n_i$, while the $\pm$ signs define $s_i$ . This
configuration represents the polymer displayed in
figure~\ref{gel.eps}. The horizontal axis labels gel pores in
sequence along the tube. The horizontal axis represents the height
$h_i$ with respect to the first pore $i=0$. In this figure L=11, S=+3,
$N_0$=7, N=19}
\end{figure}

Length and field dependence of the drift velocity have been well
studied analytically for biased reptation models~\cite{biased}. These
models assume constant tube length and conformation, independent of
the applied electric field and independent of time. They conclude that
diffusion of stored length depends only on the polymer tube length and
end-to-end distance. Their dynamics is random polymer reptation biased
by the applied electric field. Rubenstein~\cite{Rubinstein} introduced
what is now called the ``repton model'' (see figures~\ref{gel.eps}
and~\ref{repton.eps}), a lattice model that dynamically generates its
own tube. Later Duke~\cite{Duke} generalized the repton model to
include the electric field. Semenov, Duke and Viovy~\cite{BRF}
introduced fluctuations directly into the biased reptation model.

The Repton model was studied intensively by B. Widom and
co-workers~\cite{WVD,BMW,KW}, and by van Leeuwen and
co-workers~\cite{VLK,JDB,Leegwater}, among others.  It is convenient
to introduce fundamental length and frequency parameters $a$ and
$\omega$, and to define a reduced electric field $E=|q{\cal }E|a/k_B
T$.  Numerical simulations by Barkema, Marko and Widom~\cite{BMW}
suggest a simple scaling behavior for the drift velocity, with
\begin{equation}
\label{scaling}
{{N^2}\over{a \omega}} v_d(N,E) \rightarrow J(N E)
\end{equation}
in the scaling limit 
\begin{equation}
\label{scaling_limit}
E \rightarrow 0, \hspace{1cm}  N \rightarrow \infty, 
\hspace{1cm}  u \equiv N E \hspace{0.5cm} {\rm finite}
\end{equation}
They suggest an approximate functional form for the scaling function
\begin{equation}
\label{J_of_u}
J(u) \approx u
\sqrt{\left({{1}\over{3}}\right)^2+\left({{2 u}\over{5}}\right)^2}.
\end{equation}
This scaling function exhibits loss of length resolution because at
fixed fields the drift velocity varies as $E |E|$, independent of N
for long chain lengths.  Barkema, Caron and Marko~\cite{Caron}
discovered similar scaling relations for real DNA fragments and
electrophoresis gels.

The repton model decomposes the polymer chain into $N$ units called
reptons, each incorporating one persistence length of the chain.  When
neighboring reptons on the chain lie within the same gel pore a unit
of stored length separates them. Otherwise the reptons must occupy
neighboring gel pores, because the bond between reptons prevents
further excursions. Label the occupied gel pores with index $i=0, 2,
..., L$. The bond from one pore $i-1$ to the neighboring pore $i$ we
denote $s_i=\pm1$, where $i=1, 2, ...L$.  The direction of pore
labeling $i$ is chosen as follows: the ``head'' of the polymer is the
end that is furthest advanced in the direction of applied force. The
other end is the ``tail''. Labeling begins at the tail and runs to the
head. As the polymer drifts, head and tail may occasionally
interchange. The set of all bonds $\tube$ defines the shape of the
polymer's tube.

Reptons may move along the tube when attached to at least one unit
stored length. Stored length is recorded in the variable $n_i$ for the
units of stored length in gel pore $i$. The set of stored length
occupancy data for all gel pores is denoted $\occup$. The total amount
of stored length, summed over gel pores, is $N_0$=N-1-L.  When a
repton moves it carries a unit of stored length from the pore which it
left into the new pore.  The potential energy of the polymer changes
by $E s_i$ when a unit of stored length moves from pore $i-1$ to $i$.
The center of mass of the polymer advances by $s_i a/N$. Allowed moves
occur with frequency $\omega B^{s_i}$ where the Boltzmann factor
$B=e^{E/2}$. The electric field favors either left- or right-wards
motion depending on the sign of $s_i$.

The tube constantly changes shape while stored length migrates across
the polymer.  However, the changes in tube conformation for short times
occur in a small length around the ends of the polymer. The
time for stored length to traverse the tube is small compared to the
time required for significant change in polymer conformation. This is
clearly the case for zero field diffusion, because de
Gennes~\cite{deGennes} showed that the time required for tube renewal
$\tau_R$ grows as $N^3$, while the diffusion of a unit of stored
length takes the usual $\tau_S \sim N^2$ time to travel a distance N.
Thus the tube renewal time exceeds the time for stored length
relaxation by a factor N.

The factor of N, by which $\tau_R$ exceeds $\tau_S$ persists for
finite fields. Each unit of stored length that traverses the tube
advances the polymer by one gel pore. Generation of a new tube
therefore requires $L$ units of stored length to diffuse across the
polymer. Tube length $L$ is proportional to chain length N.

Examine the distance from each end of the polymer over which diffusion
of stored length is sensitive to fluctuations in tube conformation.
Denote this length scale $l_c$, and note that the time scale for
stored length diffusion across this length is $l_c^2$.  The time
needed for renewal of this length of tube varies as $l_c^3$ because
$l_c$ units of tube length must diffuse a distance of order $l_c$. The
length at which these time scales are comparable defines $l_c$, of
order $1$. Further from the ends of the tube, the tube is effectively
stationary while stored length diffuses. More information about the
coupling of stored length diffusion to chain conformation near the
chain ends may be found in reference~\cite{JDB}.

We use the idea of fixed tube conformation on the time scale of stored
length diffusion to simplify the analytic solution of the repton model
with free boundaries. By appropriately adjusting the dynamics at the
free ends, we achieve closed form expressions for the drift velocity
as a function of tube conformation. These expressions reproduce the
drift velocity of the original repton model in the limit of long
polymer chains.

\section{Stored length diffusion}

The joint probability distribution $P(\conf)$ defines the
probability for tube conformation $\tube$ and stored length
distribution $\occup$ within the tube. This function obeys the
steady-state master equation
\begin{equation}
\label{master}
\sum_{\occup' \tube'} W(\occup' \tube' \rightarrow \occup \tube)
P(\occup';\tube') =
\sum_{\occup' \tube'} W(\occup \tube \rightarrow \occup' \tube')
P(\conf)
\end{equation}
The left hand side is the rate of transitions into the state
$(\conf)$, while the right hand side is the rate of transitions out of
that state.  Approximate solution of this master equation follows the
observation that stored length diffuses quickly compared to changes in
tube conformation, except within distances of order 1 of the tube
ends. Fluctuations at the tube endpoints have negligible impact on the
stored length probability distribution away from the tube ends.

We replace the original free boundary repton model described by the
master equation~(\ref{master}) with a simpler model in which the tube
conformation $\tube$ is held fixed. The simpler master equation
\begin{eqnarray}
\label{master_tube}
\sum_{i=0}^{L} \tni \left[ \right.
B^{-s_{i+1}} 
P(\cdot \cdot \cdot ,n_i -1, n_{i+1} +1, \cdot \cdot \cdot ; \tube) 
							+ \nonumber \\
B^{s_i} P(\cdot \cdot \cdot ,n_{i-1} +1, n_i -1, \cdot \cdot \cdot ; \tube)
		\left. \right] \\
= \sum_{i=0}^{L} \tni \left[ B^{s_{i+1}} + B^{-s_i} \right] 
		P(\conf). \nonumber
\end{eqnarray}
may be solved exactly.  The first term on the left describes the flux
of stored length from gel pore $i+1$ into $i$. The second term
describes flux from $i-1$ to $i$.  The right-hand side of
equation~(\ref{master_tube}) describes flux out of site $i$. The
summations include as-yet undefined functions of stored length
occupation and bond orientation just beyond each end-point of the
tube. We define the probability function $P(\conf)$ to be independent
of the variables $n_{-1}$ and $n_{L+1}$, and we introduce fictitious
bonds $s_0=s_{L+1}=0$. This is equivalent to placing the tube in
contact with a reservoir of freely available stored length at each
end. 

Our simpler model is identical to the periodic-boundary repton model
studied by Van Leeuwen and Kooiman~\cite{VLK} with the exception of
the boundary condition. The drift velocity of sufficiently long chains
is not influenced by our alteration of the endpoint dynamics because
it is diffusion of stored length along the {\em interior} of the tube
that limits the drift velocity. The solution to
equation~(\ref{master_tube}) yields the stored length distribution
within a tube of fixed conformation $\tube$, but no information on the
probability distribution for tube conformations.
Equation~(\ref{master_tube}) has a well known  solution~\cite{derrida}
\begin{equation}
\label{master_solution}
P(\conf)=\prod_{i=0}^{L} p_i^{n_i}.
\end{equation}
Formally we extend this solution beyond the tube endpoints by defining
\begin{equation}
\label{BC}
p_{-1} \equiv p_{L+1} \equiv 1
\end{equation}
in order that $P(\conf)$ not depend on $n_{-1}$ and $n_{L+1}$.  The
form of solution~(\ref{master_solution}) with free ends is identical
to the solution with periodic boundary conditions~\cite{VLK,derrida},
because the master equation is a local difference equation.

The analysis that follows is largely adapted from the studies of the
periodic boundary repton problem, and the free boundary problem in
weak fields, by Van Leeuwen and co-workers~\cite{VLK,JDB,Leegwater}.
Our approach is exact in the large $N$ limit for the free end repton
model in arbitrary electric field.  The factors $p_i$ obey the
single-particle master equation
\begin{equation}
\label{single_particle}
B^{-s_{i+1}} p_{i+1} + B^{s_i} p_{i-1} = 
\left[ B^{s_{i+1}} + B^{-s_i} \right] p_i.
\end{equation}
The single-particle master equation~(\ref{single_particle}) is second
order, so the solution depends on two independent constants of
integration. 

One constant of integration is the flux of stored length through each
bond
\begin{equation}
\label{first_integral}
c(\tube) \equiv p_{i-1} B^{s_i} - p_i B^{-s_i}.
\end{equation}
Although the left hand side of equation~(\ref{first_integral})
formally depends on $i$, the value of $c$ is independent of $i$.  The
boundary condition $p_{-1}=1$, governing the availability of stored
length at chain ends, sets the other constant of integration. The full
solution is
\begin{equation}
\label{single_particle_solution}
p_i = p_{-1} B^{ 2 h_i } -
c \sum_{k=0}^{i} B^{ 2 (h_i - h_k) +s_k}
\end{equation}
where we define
\begin{equation}
\label{height}
h_i = \sum_{j=1}^{i} s_j 
\end{equation}
the ``height'' of the tube in the direction parallel to the field,
as indicated in figure~\ref{repton.eps}.

The form of solution~(\ref{single_particle_solution}) is interesting.
The first term is the ``barometric'' distribution of stored
length, the expected equilibrium distribution in the absence of flux.
An equivalent distribution occurs in the original repton model, when
one repton is held fixed so the polymer cannot drift~\cite{unpub}.
The second term corrects for the steady flow of particles.
Since the boundary conditions~(\ref{BC}) apply at both ends of
the chain, the stored length flux $c(\tube)$ is determined. Setting
$p_{L+1}=1$ in equation~(\ref{single_particle_solution}), we solve for
\begin{equation}
\label{flux_solution}
c(\tube)={{B^{2S}-1}\over{\sum_{k=0}^{L+1}} B^{2(S-h_k)+s_k}}.
\end{equation}
In this expression, we define
\begin{equation}
\label{S}
S = \sum_{i=1}^L s_i.
\end{equation}
as the end-to-end separation parallel to the field $E$. This quantity
equals the height at the end of the tube.

Given the probability distribution $P(\conf)$ we calculate the tube-dependent
velocity
\begin{equation}
\label{v_tube_def} 
v(\tube) = \sum_{\occup} P(\conf) v(\conf) /
		\sum_{\occup} P(\conf).
\end{equation}
Here the the imbalance of forward and backward transition rates for
stored length within the tube governs the velocity associated with a
particular allocation of stored length,
\begin{equation}
\label{v_config}
v(\conf) = {{a \omega}\over{N}} \sum_{i=0}^{L}
	\theta(n_i) (s_{i+1} B^{s_{i+1}} - s_i B^{-s_i}).
\end{equation}
We include the natural velocity scale for the problem $a \omega/N$,
with $a/N$ the center of mass displacement parallel to the field for
each repton jump, and $\omega$ the jump attempt frequency.  Once we
know the velocity for arbitrary tubes, we approximate the full drift
velocity $v_d$ as an average of $v(\tube)$ over tube conformations
$\tube$.

It is convenient to interchange the summations over $\occup$~in
equation~(\ref{v_tube_def}) with the sum over $i$ in
equation~(\ref{v_config}). Then
\begin{equation}
\label{v_tube_intermed}
v(\tube) = {{a \omega}\over{N}} \sum_{i=0}^{L} \bar{\theta_i} 
	(s_{i+1} B^{s_{i+1}} - s_i B^{s_i})
\end{equation}
where $\bar{\theta_i}$ is the average of $\theta(n_i)$ over the
allocation of stored length
\begin{equation}
\label{theta_bar_def}
\bar{\theta_i} = \sum_{\occup} P(\conf) \theta(n_i) /
	\sum_{\occup} P(\conf).
\end{equation}
Using the exact solution for $P(\conf)$
equation~(\ref{master_solution}),
\begin{equation}
\label{theta_bar_intermed}
\bar{\theta_i} = \sum_{\occup} \tni \prod_{j=0}^L p_j^{n_j} 
/ \sum_{\occup} \prod_{j=0}^L p_j^{n_j}.
\end{equation}
The theta function picks out terms that contain at least one unit of
stored length in gel pore $i$. 

Removing the obvious factor of $p_i$ from the numerator of
equation~(\ref{theta_bar_intermed}), the remaining factors 
are partition functions for the distribution of stored
length among gel pores,
\begin{equation}
\label{pf}
Q (N_s) = \sum_{\occup} \prod_{j=0}^{L} p_j^{n_j} 
\hspace{0.5cm} {\rm with} \hspace{0.5cm} \sum_{j=0}^{L} n_j = N_s.
\end{equation}
The sums in equation~(\ref{theta_bar_intermed}) allocate $N_s=N_0$
units of stored length in the denominator but only $N_s=N_0-1$ units
in the numerator. The ratio
\begin{equation}
\label{z_bar_def}
\bar{z}(\tube) \equiv Q(N_0-1)/Q(N_0)
\end{equation}
is the 
fugacity for stored length. Hence
\begin{equation}
\label{theta_bar_result}
\bar{\theta_i}=p_i \bar{z}(\tube)
\end{equation}
is the explicit solution of equation~(\ref{theta_bar_def}).

Substitute expression~(\ref{theta_bar_result}) for $\bar{\theta_i}$
into equation~(\ref{v_tube_intermed}) for $v(\tube)$. Utilizing the
fictitious bonds $s_0 = s_{L+1} = 0$ to shift the summation index on the
first term of equation~(\ref{v_tube_intermed}), yields
\begin{equation}
\label{v_tube_intermed_2}
v(\tube)={{a \omega \bar{z}(\tube)}\over{N}} 
\sum_{i=0}^L (p_{i-1} B^{s_i} - p_{i} B^{-s_i}) s_i.
\end{equation}
The quantity in parentheses above is the single particle
flux defined in equation~(\ref{first_integral}). In particular, it is
independent of $i$, and may be brought outside the sum. Finally,
we obtain our principal result
\begin{equation}
\label{v_tube_general}
v(\tube)={{a \omega S c(\tube) \bar{z}(\tube)}\over{N}}
\end{equation}
This equation is fully general, holding for any field and any tube
conformation. The relationship of $v(\tube)$ to the repton model drift
velocity $v_d$ depends on the limit of long chains. We conjecture the
average over all tubes $\tube$ of $v(\tube)$ equals the drift velocity
$v_d$ with corrections of order $1/N$ relative to $v_d$ caused by our
fixed-tube boundary conditions.

\section{Scaling limit}

In the scaling limit~(\ref{scaling_limit}) the electric field vanishes
as $1/N$, leading to significant simplifications.  For weak fields, we
simplify equation~(\ref{flux_solution}) for the single-particle flux
by expanding in powers of E
\begin{equation}
\label{weak_field_flux}
c(\tube)={{SE}\over{L+2}} + {\cal O}(E^2 S^2/L)
\end{equation}
Interestingly, this value of $c$ exactly counterbalances the
barometric term in equation~(\ref{weak_field_pi}), ensuring that the
first order term in $p_i$ grows only as EN$^{1/2}$ even if the
end-to-end distance $S$ grows proportionally to N. It inhibits the
tendency of stored length to accumulate at favorable positions on the
chain. The simple form for the flux $c$ arises because $s_i E$ is the
force along any link between gel pores, so $SE$ is the net force along
the tube, and $SE/L$ the net force per link.

Simplify equation~(\ref{single_particle_solution}) for $p_i$ by
expanding
\begin{equation}
\label{weak_field_pi}
p_i=1+E h_i -(i+1) c +{\cal O}(E^2 h_i^2).
\end{equation}
Then substitute $p_i=1+{\cal O}(E)$ into the partition functions $Q$
defined in equation~(\ref{pf}) to obtain the fugacity
\begin{equation}
\label{weak_field_fugacity}
\bar{z}(\tube)={{N_0}\over{N_0+L}}+{\cal O}(E S^2/L)
\end{equation}
with $N_0+L=N-1$.

Inserting equation~(\ref{weak_field_flux}) for flux $c(\tube)$ and
equation~(\ref{weak_field_fugacity}) for fugacity $\bar{z}(\tube)$
into equation~(\ref{v_tube_general}) for velocity, we obtain the
expression
\begin{equation}
\label{weak_field_velocity}
v(\tube)={{a \omega N_0 S^2 E}\over{N^2 L}}
\end{equation}
with corrections expected to fall off as $1/N$ in the scaling limit.
Van Leeuwen and coworkers derived similar equations for the repton
model in weak fields.  Similar expressions are also known for biased
reptation models~\cite{biased,BRF}.  We conjecture that
equation~(\ref{weak_field_velocity}) is exact for the repton model
with free boundary conditions in the scaling limit for all values of
the scaling variable $u=NE$.  We make a further conjecture, that
equation~(\ref{weak_field_velocity}) gives the drift velocity $v_d$
with $N_0$, $L$ and $S$ set to their average values.

One immediate result of expression~(\ref{weak_field_velocity}) is the
zero-field diffusion constant. Since $S^2=L=2N/3$ at zero field, the
weak-field drift velocity is simply $a \omega (NE)/3N^2$, in agreement
with the accepted diffusion constant $D=1/3N^2$. Hence our conjectures
hold at least at $u=0$.

\section{Numerical study and discussion}

To test our conjecture that equation (\ref{weak_field_velocity}) holds
throughout the scaling limit~(\ref{scaling_limit}) we calculate
averages of L, $S^2$ and $v(\tube)/v_d$ for various values of the
scaling variable $u=N E$, and then extrapolate to large chain lengths
N.  We test the conjecture for moderate and large $u=1$ and $10$. For
short chain lengths, N=3-7 we perform exact calculations based on the
matrix approach described in reference~\cite{WVD}. Because the matrix
dimensionality grows as $3^N$ it is impractical to extend these
calculations to large N.  Instead, we perform numerical simulation of
drift using the multi-spin simulation program of
Gerard~Barkema~\cite{BMW}.  During these simulations we monitor $S^2$
and $L$ in addition to the drift velocity.

\begin{figure}
% GNUPLOT: LaTeX picture
\setlength{\unitlength}{0.240900pt}
\ifx\plotpoint\undefined\newsavebox{\plotpoint}\fi
\begin{picture}(1500,900)(0,0)
\font\gnuplot=cmr10 at 10pt
\gnuplot
\sbox{\plotpoint}{\rule[-0.200pt]{0.400pt}{0.400pt}}%
\put(241.0,134.0){\rule[-0.200pt]{4.818pt}{0.400pt}}
\put(219,134){\makebox(0,0)[r]{0.666}}
\put(1416.0,134.0){\rule[-0.200pt]{4.818pt}{0.400pt}}
\put(241.0,314.0){\rule[-0.200pt]{4.818pt}{0.400pt}}
\put(219,314){\makebox(0,0)[r]{0.668}}
\put(1416.0,314.0){\rule[-0.200pt]{4.818pt}{0.400pt}}
\put(241.0,494.0){\rule[-0.200pt]{4.818pt}{0.400pt}}
\put(219,494){\makebox(0,0)[r]{0.67}}
\put(1416.0,494.0){\rule[-0.200pt]{4.818pt}{0.400pt}}
\put(241.0,675.0){\rule[-0.200pt]{4.818pt}{0.400pt}}
\put(219,675){\makebox(0,0)[r]{0.672}}
\put(1416.0,675.0){\rule[-0.200pt]{4.818pt}{0.400pt}}
\put(241.0,855.0){\rule[-0.200pt]{4.818pt}{0.400pt}}
\put(219,855){\makebox(0,0)[r]{0.674}}
\put(1416.0,855.0){\rule[-0.200pt]{4.818pt}{0.400pt}}
\put(241.0,134.0){\rule[-0.200pt]{0.400pt}{4.818pt}}
\put(241,89){\makebox(0,0){0}}
\put(241.0,835.0){\rule[-0.200pt]{0.400pt}{4.818pt}}
\put(480.0,134.0){\rule[-0.200pt]{0.400pt}{4.818pt}}
\put(480,89){\makebox(0,0){50}}
\put(480.0,835.0){\rule[-0.200pt]{0.400pt}{4.818pt}}
\put(719.0,134.0){\rule[-0.200pt]{0.400pt}{4.818pt}}
\put(719,89){\makebox(0,0){100}}
\put(719.0,835.0){\rule[-0.200pt]{0.400pt}{4.818pt}}
\put(958.0,134.0){\rule[-0.200pt]{0.400pt}{4.818pt}}
\put(958,89){\makebox(0,0){150}}
\put(958.0,835.0){\rule[-0.200pt]{0.400pt}{4.818pt}}
\put(1197.0,134.0){\rule[-0.200pt]{0.400pt}{4.818pt}}
\put(1197,89){\makebox(0,0){200}}
\put(1197.0,835.0){\rule[-0.200pt]{0.400pt}{4.818pt}}
\put(1436.0,134.0){\rule[-0.200pt]{0.400pt}{4.818pt}}
\put(1436,89){\makebox(0,0){250}}
\put(1436.0,835.0){\rule[-0.200pt]{0.400pt}{4.818pt}}
\put(241.0,134.0){\rule[-0.200pt]{287.875pt}{0.400pt}}
\put(1436.0,134.0){\rule[-0.200pt]{0.400pt}{173.689pt}}
\put(241.0,855.0){\rule[-0.200pt]{287.875pt}{0.400pt}}
\put(45,494){\makebox(0,0){${\overline{L}}\over{N-1}$}}
\put(838,44){\makebox(0,0){N}}
\put(241.0,134.0){\rule[-0.200pt]{0.400pt}{173.689pt}}
\put(255,745){\raisebox{-.8pt}{\makebox(0,0){$\Diamond$}}}
\put(260,697){\raisebox{-.8pt}{\makebox(0,0){$\Diamond$}}}
\put(265,691){\raisebox{-.8pt}{\makebox(0,0){$\Diamond$}}}
\put(270,681){\raisebox{-.8pt}{\makebox(0,0){$\Diamond$}}}
\put(274,671){\raisebox{-.8pt}{\makebox(0,0){$\Diamond$}}}
\put(480,492){\raisebox{-.8pt}{\makebox(0,0){$\Diamond$}}}
\put(719,405){\raisebox{-.8pt}{\makebox(0,0){$\Diamond$}}}
\put(958,313){\raisebox{-.8pt}{\makebox(0,0){$\Diamond$}}}
\put(1197,309){\raisebox{-.8pt}{\makebox(0,0){$\Diamond$}}}
\put(241,194){\usebox{\plotpoint}}
\put(241.00,194.00){\usebox{\plotpoint}}
\put(261.76,194.00){\usebox{\plotpoint}}
\put(282.51,194.00){\usebox{\plotpoint}}
\put(303.27,194.00){\usebox{\plotpoint}}
\put(324.02,194.00){\usebox{\plotpoint}}
\put(344.78,194.00){\usebox{\plotpoint}}
\put(365.53,194.00){\usebox{\plotpoint}}
\put(386.29,194.00){\usebox{\plotpoint}}
\put(407.04,194.00){\usebox{\plotpoint}}
\put(427.80,194.00){\usebox{\plotpoint}}
\put(448.55,194.00){\usebox{\plotpoint}}
\put(469.31,194.00){\usebox{\plotpoint}}
\put(490.07,194.00){\usebox{\plotpoint}}
\put(510.82,194.00){\usebox{\plotpoint}}
\put(531.58,194.00){\usebox{\plotpoint}}
\put(552.33,194.00){\usebox{\plotpoint}}
\put(573.09,194.00){\usebox{\plotpoint}}
\put(593.84,194.00){\usebox{\plotpoint}}
\put(614.60,194.00){\usebox{\plotpoint}}
\put(635.35,194.00){\usebox{\plotpoint}}
\put(656.11,194.00){\usebox{\plotpoint}}
\put(676.87,194.00){\usebox{\plotpoint}}
\put(697.62,194.00){\usebox{\plotpoint}}
\put(718.38,194.00){\usebox{\plotpoint}}
\put(739.13,194.00){\usebox{\plotpoint}}
\put(759.89,194.00){\usebox{\plotpoint}}
\put(780.64,194.00){\usebox{\plotpoint}}
\put(801.40,194.00){\usebox{\plotpoint}}
\put(822.15,194.00){\usebox{\plotpoint}}
\put(842.91,194.00){\usebox{\plotpoint}}
\put(863.66,194.00){\usebox{\plotpoint}}
\put(884.42,194.00){\usebox{\plotpoint}}
\put(905.18,194.00){\usebox{\plotpoint}}
\put(925.93,194.00){\usebox{\plotpoint}}
\put(946.69,194.00){\usebox{\plotpoint}}
\put(967.44,194.00){\usebox{\plotpoint}}
\put(988.20,194.00){\usebox{\plotpoint}}
\put(1008.95,194.00){\usebox{\plotpoint}}
\put(1029.71,194.00){\usebox{\plotpoint}}
\put(1050.46,194.00){\usebox{\plotpoint}}
\put(1071.22,194.00){\usebox{\plotpoint}}
\put(1091.98,194.00){\usebox{\plotpoint}}
\put(1112.73,194.00){\usebox{\plotpoint}}
\put(1133.49,194.00){\usebox{\plotpoint}}
\put(1154.24,194.00){\usebox{\plotpoint}}
\put(1175.00,194.00){\usebox{\plotpoint}}
\put(1195.75,194.00){\usebox{\plotpoint}}
\put(1216.51,194.00){\usebox{\plotpoint}}
\put(1237.26,194.00){\usebox{\plotpoint}}
\put(1258.02,194.00){\usebox{\plotpoint}}
\put(1278.77,194.00){\usebox{\plotpoint}}
\put(1299.53,194.00){\usebox{\plotpoint}}
\put(1320.29,194.00){\usebox{\plotpoint}}
\put(1341.04,194.00){\usebox{\plotpoint}}
\put(1361.80,194.00){\usebox{\plotpoint}}
\put(1382.55,194.00){\usebox{\plotpoint}}
\put(1403.31,194.00){\usebox{\plotpoint}}
\put(1424.06,194.00){\usebox{\plotpoint}}
\put(1436,194){\usebox{\plotpoint}}
\end{picture}
% GNUPLOT: LaTeX picture
\setlength{\unitlength}{0.240900pt}
\ifx\plotpoint\undefined\newsavebox{\plotpoint}\fi
\begin{picture}(1500,900)(0,0)
\font\gnuplot=cmr10 at 10pt
\gnuplot
\sbox{\plotpoint}{\rule[-0.200pt]{0.400pt}{0.400pt}}%
\put(197.0,134.0){\rule[-0.200pt]{4.818pt}{0.400pt}}
\put(175,134){\makebox(0,0)[r]{0.6}}
\put(1416.0,134.0){\rule[-0.200pt]{4.818pt}{0.400pt}}
\put(197.0,294.0){\rule[-0.200pt]{4.818pt}{0.400pt}}
\put(175,294){\makebox(0,0)[r]{0.7}}
\put(1416.0,294.0){\rule[-0.200pt]{4.818pt}{0.400pt}}
\put(197.0,454.0){\rule[-0.200pt]{4.818pt}{0.400pt}}
\put(175,454){\makebox(0,0)[r]{0.8}}
\put(1416.0,454.0){\rule[-0.200pt]{4.818pt}{0.400pt}}
\put(197.0,615.0){\rule[-0.200pt]{4.818pt}{0.400pt}}
\put(175,615){\makebox(0,0)[r]{0.9}}
\put(1416.0,615.0){\rule[-0.200pt]{4.818pt}{0.400pt}}
\put(197.0,775.0){\rule[-0.200pt]{4.818pt}{0.400pt}}
\put(175,775){\makebox(0,0)[r]{1}}
\put(1416.0,775.0){\rule[-0.200pt]{4.818pt}{0.400pt}}
\put(197.0,134.0){\rule[-0.200pt]{0.400pt}{4.818pt}}
\put(197,89){\makebox(0,0){0}}
\put(197.0,835.0){\rule[-0.200pt]{0.400pt}{4.818pt}}
\put(445.0,134.0){\rule[-0.200pt]{0.400pt}{4.818pt}}
\put(445,89){\makebox(0,0){50}}
\put(445.0,835.0){\rule[-0.200pt]{0.400pt}{4.818pt}}
\put(693.0,134.0){\rule[-0.200pt]{0.400pt}{4.818pt}}
\put(693,89){\makebox(0,0){100}}
\put(693.0,835.0){\rule[-0.200pt]{0.400pt}{4.818pt}}
\put(940.0,134.0){\rule[-0.200pt]{0.400pt}{4.818pt}}
\put(940,89){\makebox(0,0){150}}
\put(940.0,835.0){\rule[-0.200pt]{0.400pt}{4.818pt}}
\put(1188.0,134.0){\rule[-0.200pt]{0.400pt}{4.818pt}}
\put(1188,89){\makebox(0,0){200}}
\put(1188.0,835.0){\rule[-0.200pt]{0.400pt}{4.818pt}}
\put(1436.0,134.0){\rule[-0.200pt]{0.400pt}{4.818pt}}
\put(1436,89){\makebox(0,0){250}}
\put(1436.0,835.0){\rule[-0.200pt]{0.400pt}{4.818pt}}
\put(197.0,134.0){\rule[-0.200pt]{298.475pt}{0.400pt}}
\put(1436.0,134.0){\rule[-0.200pt]{0.400pt}{173.689pt}}
\put(197.0,855.0){\rule[-0.200pt]{298.475pt}{0.400pt}}
\put(45,494){\makebox(0,0){${\overline{S^2}}\over{(N-1)}$}}
\put(816,44){\makebox(0,0){N}}
\put(197.0,134.0){\rule[-0.200pt]{0.400pt}{173.689pt}}
\put(212,216){\raisebox{-.8pt}{\makebox(0,0){$\Diamond$}}}
\put(217,231){\raisebox{-.8pt}{\makebox(0,0){$\Diamond$}}}
\put(222,241){\raisebox{-.8pt}{\makebox(0,0){$\Diamond$}}}
\put(227,249){\raisebox{-.8pt}{\makebox(0,0){$\Diamond$}}}
\put(232,258){\raisebox{-.8pt}{\makebox(0,0){$\Diamond$}}}
\put(445,512){\raisebox{-.8pt}{\makebox(0,0){$\Diamond$}}}
\put(693,629){\raisebox{-.8pt}{\makebox(0,0){$\Diamond$}}}
\put(940,697){\raisebox{-.8pt}{\makebox(0,0){$\Diamond$}}}
\put(1188,750){\raisebox{-.8pt}{\makebox(0,0){$\Diamond$}}}
\end{picture}
% GNUPLOT: LaTeX picture
\setlength{\unitlength}{0.240900pt}
\ifx\plotpoint\undefined\newsavebox{\plotpoint}\fi
\begin{picture}(1500,900)(0,0)
\font\gnuplot=cmr10 at 10pt
\gnuplot
\sbox{\plotpoint}{\rule[-0.200pt]{0.400pt}{0.400pt}}%
\put(197.0,134.0){\rule[-0.200pt]{4.818pt}{0.400pt}}
\put(175,134){\makebox(0,0)[r]{0}}
\put(1416.0,134.0){\rule[-0.200pt]{4.818pt}{0.400pt}}
\put(197.0,278.0){\rule[-0.200pt]{4.818pt}{0.400pt}}
\put(175,278){\makebox(0,0)[r]{0.2}}
\put(1416.0,278.0){\rule[-0.200pt]{4.818pt}{0.400pt}}
\put(197.0,422.0){\rule[-0.200pt]{4.818pt}{0.400pt}}
\put(175,422){\makebox(0,0)[r]{0.4}}
\put(1416.0,422.0){\rule[-0.200pt]{4.818pt}{0.400pt}}
\put(197.0,567.0){\rule[-0.200pt]{4.818pt}{0.400pt}}
\put(175,567){\makebox(0,0)[r]{0.6}}
\put(1416.0,567.0){\rule[-0.200pt]{4.818pt}{0.400pt}}
\put(197.0,711.0){\rule[-0.200pt]{4.818pt}{0.400pt}}
\put(175,711){\makebox(0,0)[r]{0.8}}
\put(1416.0,711.0){\rule[-0.200pt]{4.818pt}{0.400pt}}
\put(197.0,855.0){\rule[-0.200pt]{4.818pt}{0.400pt}}
\put(175,855){\makebox(0,0)[r]{1}}
\put(1416.0,855.0){\rule[-0.200pt]{4.818pt}{0.400pt}}
\put(197.0,134.0){\rule[-0.200pt]{0.400pt}{4.818pt}}
\put(197,89){\makebox(0,0){0}}
\put(197.0,835.0){\rule[-0.200pt]{0.400pt}{4.818pt}}
\put(445.0,134.0){\rule[-0.200pt]{0.400pt}{4.818pt}}
\put(445,89){\makebox(0,0){50}}
\put(445.0,835.0){\rule[-0.200pt]{0.400pt}{4.818pt}}
\put(693.0,134.0){\rule[-0.200pt]{0.400pt}{4.818pt}}
\put(693,89){\makebox(0,0){100}}
\put(693.0,835.0){\rule[-0.200pt]{0.400pt}{4.818pt}}
\put(940.0,134.0){\rule[-0.200pt]{0.400pt}{4.818pt}}
\put(940,89){\makebox(0,0){150}}
\put(940.0,835.0){\rule[-0.200pt]{0.400pt}{4.818pt}}
\put(1188.0,134.0){\rule[-0.200pt]{0.400pt}{4.818pt}}
\put(1188,89){\makebox(0,0){200}}
\put(1188.0,835.0){\rule[-0.200pt]{0.400pt}{4.818pt}}
\put(1436.0,134.0){\rule[-0.200pt]{0.400pt}{4.818pt}}
\put(1436,89){\makebox(0,0){250}}
\put(1436.0,835.0){\rule[-0.200pt]{0.400pt}{4.818pt}}
\put(197.0,134.0){\rule[-0.200pt]{298.475pt}{0.400pt}}
\put(1436.0,134.0){\rule[-0.200pt]{0.400pt}{173.689pt}}
\put(197.0,855.0){\rule[-0.200pt]{298.475pt}{0.400pt}}
\put(45,494){\makebox(0,0){${v({\bf s})}\over{v_d}$}}
\put(816,44){\makebox(0,0){N}}
\put(197.0,134.0){\rule[-0.200pt]{0.400pt}{173.689pt}}
\put(212,288){\raisebox{-.8pt}{\makebox(0,0){$\Diamond$}}}
\put(217,325){\raisebox{-.8pt}{\makebox(0,0){$\Diamond$}}}
\put(222,354){\raisebox{-.8pt}{\makebox(0,0){$\Diamond$}}}
\put(227,380){\raisebox{-.8pt}{\makebox(0,0){$\Diamond$}}}
\put(232,402){\raisebox{-.8pt}{\makebox(0,0){$\Diamond$}}}
\put(445,656){\raisebox{-.8pt}{\makebox(0,0){$\Diamond$}}}
\put(693,711){\raisebox{-.8pt}{\makebox(0,0){$\Diamond$}}}
\put(940,743){\raisebox{-.8pt}{\makebox(0,0){$\Diamond$}}}
\put(1188,769){\raisebox{-.8pt}{\makebox(0,0){$\Diamond$}}}
\end{picture}
\caption{
\label{u_one}
Moderate field $NE=1$ data. $L/(N-1)$, $S^2/(N-1)$ and $v(\tube)/v_d$
are shown, respectively, in parts a, b and c.}
\end{figure}

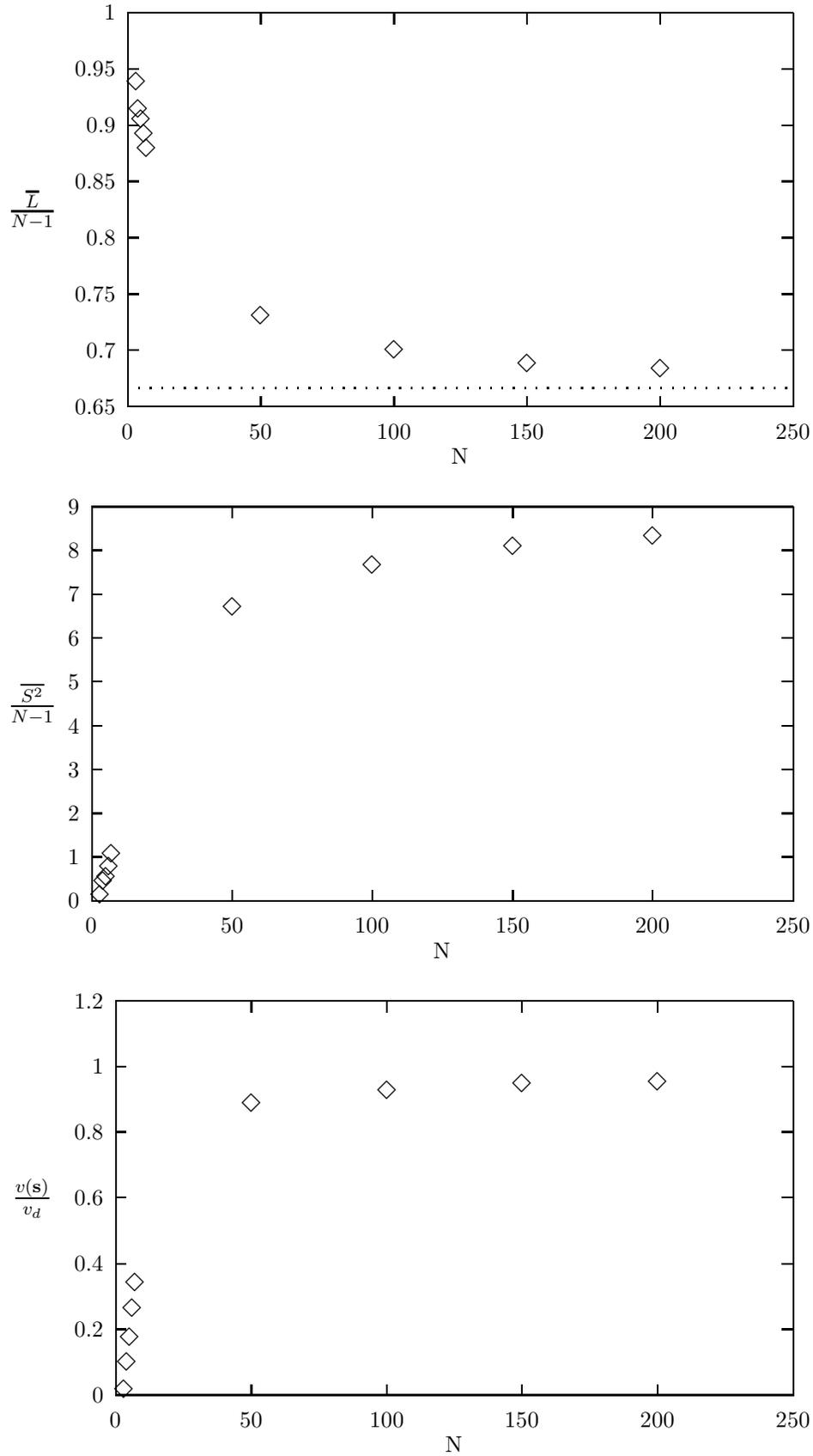
\begin{figure}
% GNUPLOT: LaTeX picture
\setlength{\unitlength}{0.240900pt}
\ifx\plotpoint\undefined\newsavebox{\plotpoint}\fi
\begin{picture}(1500,900)(0,0)
\font\gnuplot=cmr10 at 10pt
\gnuplot
\sbox{\plotpoint}{\rule[-0.200pt]{0.400pt}{0.400pt}}%
\put(219.0,134.0){\rule[-0.200pt]{4.818pt}{0.400pt}}
\put(197,134){\makebox(0,0)[r]{0.65}}
\put(1416.0,134.0){\rule[-0.200pt]{4.818pt}{0.400pt}}
\put(219.0,237.0){\rule[-0.200pt]{4.818pt}{0.400pt}}
\put(197,237){\makebox(0,0)[r]{0.7}}
\put(1416.0,237.0){\rule[-0.200pt]{4.818pt}{0.400pt}}
\put(219.0,340.0){\rule[-0.200pt]{4.818pt}{0.400pt}}
\put(197,340){\makebox(0,0)[r]{0.75}}
\put(1416.0,340.0){\rule[-0.200pt]{4.818pt}{0.400pt}}
\put(219.0,443.0){\rule[-0.200pt]{4.818pt}{0.400pt}}
\put(197,443){\makebox(0,0)[r]{0.8}}
\put(1416.0,443.0){\rule[-0.200pt]{4.818pt}{0.400pt}}
\put(219.0,546.0){\rule[-0.200pt]{4.818pt}{0.400pt}}
\put(197,546){\makebox(0,0)[r]{0.85}}
\put(1416.0,546.0){\rule[-0.200pt]{4.818pt}{0.400pt}}
\put(219.0,649.0){\rule[-0.200pt]{4.818pt}{0.400pt}}
\put(197,649){\makebox(0,0)[r]{0.9}}
\put(1416.0,649.0){\rule[-0.200pt]{4.818pt}{0.400pt}}
\put(219.0,752.0){\rule[-0.200pt]{4.818pt}{0.400pt}}
\put(197,752){\makebox(0,0)[r]{0.95}}
\put(1416.0,752.0){\rule[-0.200pt]{4.818pt}{0.400pt}}
\put(219.0,855.0){\rule[-0.200pt]{4.818pt}{0.400pt}}
\put(197,855){\makebox(0,0)[r]{1}}
\put(1416.0,855.0){\rule[-0.200pt]{4.818pt}{0.400pt}}
\put(219.0,134.0){\rule[-0.200pt]{0.400pt}{4.818pt}}
\put(219,89){\makebox(0,0){0}}
\put(219.0,835.0){\rule[-0.200pt]{0.400pt}{4.818pt}}
\put(462.0,134.0){\rule[-0.200pt]{0.400pt}{4.818pt}}
\put(462,89){\makebox(0,0){50}}
\put(462.0,835.0){\rule[-0.200pt]{0.400pt}{4.818pt}}
\put(706.0,134.0){\rule[-0.200pt]{0.400pt}{4.818pt}}
\put(706,89){\makebox(0,0){100}}
\put(706.0,835.0){\rule[-0.200pt]{0.400pt}{4.818pt}}
\put(949.0,134.0){\rule[-0.200pt]{0.400pt}{4.818pt}}
\put(949,89){\makebox(0,0){150}}
\put(949.0,835.0){\rule[-0.200pt]{0.400pt}{4.818pt}}
\put(1193.0,134.0){\rule[-0.200pt]{0.400pt}{4.818pt}}
\put(1193,89){\makebox(0,0){200}}
\put(1193.0,835.0){\rule[-0.200pt]{0.400pt}{4.818pt}}
\put(1436.0,134.0){\rule[-0.200pt]{0.400pt}{4.818pt}}
\put(1436,89){\makebox(0,0){250}}
\put(1436.0,835.0){\rule[-0.200pt]{0.400pt}{4.818pt}}
\put(219.0,134.0){\rule[-0.200pt]{293.175pt}{0.400pt}}
\put(1436.0,134.0){\rule[-0.200pt]{0.400pt}{173.689pt}}
\put(219.0,855.0){\rule[-0.200pt]{293.175pt}{0.400pt}}
\put(45,494){\makebox(0,0){${\overline{L}}\over{N-1}$}}
\put(827,44){\makebox(0,0){N}}
\put(219.0,134.0){\rule[-0.200pt]{0.400pt}{173.689pt}}
\put(234,727){\raisebox{-.8pt}{\makebox(0,0){$\Diamond$}}}
\put(238,677){\raisebox{-.8pt}{\makebox(0,0){$\Diamond$}}}
\put(243,658){\raisebox{-.8pt}{\makebox(0,0){$\Diamond$}}}
\put(248,632){\raisebox{-.8pt}{\makebox(0,0){$\Diamond$}}}
\put(253,605){\raisebox{-.8pt}{\makebox(0,0){$\Diamond$}}}
\put(462,299){\raisebox{-.8pt}{\makebox(0,0){$\Diamond$}}}
\put(706,236){\raisebox{-.8pt}{\makebox(0,0){$\Diamond$}}}
\put(949,212){\raisebox{-.8pt}{\makebox(0,0){$\Diamond$}}}
\put(1193,202){\raisebox{-.8pt}{\makebox(0,0){$\Diamond$}}}
\put(219,168){\usebox{\plotpoint}}
\put(219.00,168.00){\usebox{\plotpoint}}
\put(239.76,168.00){\usebox{\plotpoint}}
\put(260.51,168.00){\usebox{\plotpoint}}
\put(281.27,168.00){\usebox{\plotpoint}}
\put(302.02,168.00){\usebox{\plotpoint}}
\put(322.78,168.00){\usebox{\plotpoint}}
\put(343.53,168.00){\usebox{\plotpoint}}
\put(364.29,168.00){\usebox{\plotpoint}}
\put(385.04,168.00){\usebox{\plotpoint}}
\put(405.80,168.00){\usebox{\plotpoint}}
\put(426.55,168.00){\usebox{\plotpoint}}
\put(447.31,168.00){\usebox{\plotpoint}}
\put(468.07,168.00){\usebox{\plotpoint}}
\put(488.82,168.00){\usebox{\plotpoint}}
\put(509.58,168.00){\usebox{\plotpoint}}
\put(530.33,168.00){\usebox{\plotpoint}}
\put(551.09,168.00){\usebox{\plotpoint}}
\put(571.84,168.00){\usebox{\plotpoint}}
\put(592.60,168.00){\usebox{\plotpoint}}
\put(613.35,168.00){\usebox{\plotpoint}}
\put(634.11,168.00){\usebox{\plotpoint}}
\put(654.87,168.00){\usebox{\plotpoint}}
\put(675.62,168.00){\usebox{\plotpoint}}
\put(696.38,168.00){\usebox{\plotpoint}}
\put(717.13,168.00){\usebox{\plotpoint}}
\put(737.89,168.00){\usebox{\plotpoint}}
\put(758.64,168.00){\usebox{\plotpoint}}
\put(779.40,168.00){\usebox{\plotpoint}}
\put(800.15,168.00){\usebox{\plotpoint}}
\put(820.91,168.00){\usebox{\plotpoint}}
\put(841.66,168.00){\usebox{\plotpoint}}
\put(862.42,168.00){\usebox{\plotpoint}}
\put(883.18,168.00){\usebox{\plotpoint}}
\put(903.93,168.00){\usebox{\plotpoint}}
\put(924.69,168.00){\usebox{\plotpoint}}
\put(945.44,168.00){\usebox{\plotpoint}}
\put(966.20,168.00){\usebox{\plotpoint}}
\put(986.95,168.00){\usebox{\plotpoint}}
\put(1007.71,168.00){\usebox{\plotpoint}}
\put(1028.46,168.00){\usebox{\plotpoint}}
\put(1049.22,168.00){\usebox{\plotpoint}}
\put(1069.98,168.00){\usebox{\plotpoint}}
\put(1090.73,168.00){\usebox{\plotpoint}}
\put(1111.49,168.00){\usebox{\plotpoint}}
\put(1132.24,168.00){\usebox{\plotpoint}}
\put(1153.00,168.00){\usebox{\plotpoint}}
\put(1173.75,168.00){\usebox{\plotpoint}}
\put(1194.51,168.00){\usebox{\plotpoint}}
\put(1215.26,168.00){\usebox{\plotpoint}}
\put(1236.02,168.00){\usebox{\plotpoint}}
\put(1256.77,168.00){\usebox{\plotpoint}}
\put(1277.53,168.00){\usebox{\plotpoint}}
\put(1298.29,168.00){\usebox{\plotpoint}}
\put(1319.04,168.00){\usebox{\plotpoint}}
\put(1339.80,168.00){\usebox{\plotpoint}}
\put(1360.55,168.00){\usebox{\plotpoint}}
\put(1381.31,168.00){\usebox{\plotpoint}}
\put(1402.06,168.00){\usebox{\plotpoint}}
\put(1422.82,168.00){\usebox{\plotpoint}}
\put(1436,168){\usebox{\plotpoint}}
\end{picture}
% GNUPLOT: LaTeX picture
\setlength{\unitlength}{0.240900pt}
\ifx\plotpoint\undefined\newsavebox{\plotpoint}\fi
\begin{picture}(1500,900)(0,0)
\font\gnuplot=cmr10 at 10pt
\gnuplot
\sbox{\plotpoint}{\rule[-0.200pt]{0.400pt}{0.400pt}}%
\put(153.0,134.0){\rule[-0.200pt]{4.818pt}{0.400pt}}
\put(131,134){\makebox(0,0)[r]{0}}
\put(1416.0,134.0){\rule[-0.200pt]{4.818pt}{0.400pt}}
\put(153.0,214.0){\rule[-0.200pt]{4.818pt}{0.400pt}}
\put(131,214){\makebox(0,0)[r]{1}}
\put(1416.0,214.0){\rule[-0.200pt]{4.818pt}{0.400pt}}
\put(153.0,294.0){\rule[-0.200pt]{4.818pt}{0.400pt}}
\put(131,294){\makebox(0,0)[r]{2}}
\put(1416.0,294.0){\rule[-0.200pt]{4.818pt}{0.400pt}}
\put(153.0,374.0){\rule[-0.200pt]{4.818pt}{0.400pt}}
\put(131,374){\makebox(0,0)[r]{3}}
\put(1416.0,374.0){\rule[-0.200pt]{4.818pt}{0.400pt}}
\put(153.0,454.0){\rule[-0.200pt]{4.818pt}{0.400pt}}
\put(131,454){\makebox(0,0)[r]{4}}
\put(1416.0,454.0){\rule[-0.200pt]{4.818pt}{0.400pt}}
\put(153.0,535.0){\rule[-0.200pt]{4.818pt}{0.400pt}}
\put(131,535){\makebox(0,0)[r]{5}}
\put(1416.0,535.0){\rule[-0.200pt]{4.818pt}{0.400pt}}
\put(153.0,615.0){\rule[-0.200pt]{4.818pt}{0.400pt}}
\put(131,615){\makebox(0,0)[r]{6}}
\put(1416.0,615.0){\rule[-0.200pt]{4.818pt}{0.400pt}}
\put(153.0,695.0){\rule[-0.200pt]{4.818pt}{0.400pt}}
\put(131,695){\makebox(0,0)[r]{7}}
\put(1416.0,695.0){\rule[-0.200pt]{4.818pt}{0.400pt}}
\put(153.0,775.0){\rule[-0.200pt]{4.818pt}{0.400pt}}
\put(131,775){\makebox(0,0)[r]{8}}
\put(1416.0,775.0){\rule[-0.200pt]{4.818pt}{0.400pt}}
\put(153.0,855.0){\rule[-0.200pt]{4.818pt}{0.400pt}}
\put(131,855){\makebox(0,0)[r]{9}}
\put(1416.0,855.0){\rule[-0.200pt]{4.818pt}{0.400pt}}
\put(153.0,134.0){\rule[-0.200pt]{0.400pt}{4.818pt}}
\put(153,89){\makebox(0,0){0}}
\put(153.0,835.0){\rule[-0.200pt]{0.400pt}{4.818pt}}
\put(410.0,134.0){\rule[-0.200pt]{0.400pt}{4.818pt}}
\put(410,89){\makebox(0,0){50}}
\put(410.0,835.0){\rule[-0.200pt]{0.400pt}{4.818pt}}
\put(666.0,134.0){\rule[-0.200pt]{0.400pt}{4.818pt}}
\put(666,89){\makebox(0,0){100}}
\put(666.0,835.0){\rule[-0.200pt]{0.400pt}{4.818pt}}
\put(923.0,134.0){\rule[-0.200pt]{0.400pt}{4.818pt}}
\put(923,89){\makebox(0,0){150}}
\put(923.0,835.0){\rule[-0.200pt]{0.400pt}{4.818pt}}
\put(1179.0,134.0){\rule[-0.200pt]{0.400pt}{4.818pt}}
\put(1179,89){\makebox(0,0){200}}
\put(1179.0,835.0){\rule[-0.200pt]{0.400pt}{4.818pt}}
\put(1436.0,134.0){\rule[-0.200pt]{0.400pt}{4.818pt}}
\put(1436,89){\makebox(0,0){250}}
\put(1436.0,835.0){\rule[-0.200pt]{0.400pt}{4.818pt}}
\put(153.0,134.0){\rule[-0.200pt]{309.075pt}{0.400pt}}
\put(1436.0,134.0){\rule[-0.200pt]{0.400pt}{173.689pt}}
\put(153.0,855.0){\rule[-0.200pt]{309.075pt}{0.400pt}}
\put(45,494){\makebox(0,0){${\overline{S^2}}\over{N-1}$}}
\put(794,44){\makebox(0,0){N}}
\put(153.0,134.0){\rule[-0.200pt]{0.400pt}{173.689pt}}
\put(168,143){\raisebox{-.8pt}{\makebox(0,0){$\Diamond$}}}
\put(174,168){\raisebox{-.8pt}{\makebox(0,0){$\Diamond$}}}
\put(179,176){\raisebox{-.8pt}{\makebox(0,0){$\Diamond$}}}
\put(184,195){\raisebox{-.8pt}{\makebox(0,0){$\Diamond$}}}
\put(189,219){\raisebox{-.8pt}{\makebox(0,0){$\Diamond$}}}
\put(410,670){\raisebox{-.8pt}{\makebox(0,0){$\Diamond$}}}
\put(666,746){\raisebox{-.8pt}{\makebox(0,0){$\Diamond$}}}
\put(923,781){\raisebox{-.8pt}{\makebox(0,0){$\Diamond$}}}
\put(1179,800){\raisebox{-.8pt}{\makebox(0,0){$\Diamond$}}}
\end{picture}
% GNUPLOT: LaTeX picture
\setlength{\unitlength}{0.240900pt}
\ifx\plotpoint\undefined\newsavebox{\plotpoint}\fi
\begin{picture}(1500,900)(0,0)
\font\gnuplot=cmr10 at 10pt
\gnuplot
\sbox{\plotpoint}{\rule[-0.200pt]{0.400pt}{0.400pt}}%
\put(197.0,134.0){\rule[-0.200pt]{4.818pt}{0.400pt}}
\put(175,134){\makebox(0,0)[r]{0}}
\put(1416.0,134.0){\rule[-0.200pt]{4.818pt}{0.400pt}}
\put(197.0,254.0){\rule[-0.200pt]{4.818pt}{0.400pt}}
\put(175,254){\makebox(0,0)[r]{0.2}}
\put(1416.0,254.0){\rule[-0.200pt]{4.818pt}{0.400pt}}
\put(197.0,374.0){\rule[-0.200pt]{4.818pt}{0.400pt}}
\put(175,374){\makebox(0,0)[r]{0.4}}
\put(1416.0,374.0){\rule[-0.200pt]{4.818pt}{0.400pt}}
\put(197.0,495.0){\rule[-0.200pt]{4.818pt}{0.400pt}}
\put(175,495){\makebox(0,0)[r]{0.6}}
\put(1416.0,495.0){\rule[-0.200pt]{4.818pt}{0.400pt}}
\put(197.0,615.0){\rule[-0.200pt]{4.818pt}{0.400pt}}
\put(175,615){\makebox(0,0)[r]{0.8}}
\put(1416.0,615.0){\rule[-0.200pt]{4.818pt}{0.400pt}}
\put(197.0,735.0){\rule[-0.200pt]{4.818pt}{0.400pt}}
\put(175,735){\makebox(0,0)[r]{1}}
\put(1416.0,735.0){\rule[-0.200pt]{4.818pt}{0.400pt}}
\put(197.0,855.0){\rule[-0.200pt]{4.818pt}{0.400pt}}
\put(175,855){\makebox(0,0)[r]{1.2}}
\put(1416.0,855.0){\rule[-0.200pt]{4.818pt}{0.400pt}}
\put(197.0,134.0){\rule[-0.200pt]{0.400pt}{4.818pt}}
\put(197,89){\makebox(0,0){0}}
\put(197.0,835.0){\rule[-0.200pt]{0.400pt}{4.818pt}}
\put(445.0,134.0){\rule[-0.200pt]{0.400pt}{4.818pt}}
\put(445,89){\makebox(0,0){50}}
\put(445.0,835.0){\rule[-0.200pt]{0.400pt}{4.818pt}}
\put(693.0,134.0){\rule[-0.200pt]{0.400pt}{4.818pt}}
\put(693,89){\makebox(0,0){100}}
\put(693.0,835.0){\rule[-0.200pt]{0.400pt}{4.818pt}}
\put(940.0,134.0){\rule[-0.200pt]{0.400pt}{4.818pt}}
\put(940,89){\makebox(0,0){150}}
\put(940.0,835.0){\rule[-0.200pt]{0.400pt}{4.818pt}}
\put(1188.0,134.0){\rule[-0.200pt]{0.400pt}{4.818pt}}
\put(1188,89){\makebox(0,0){200}}
\put(1188.0,835.0){\rule[-0.200pt]{0.400pt}{4.818pt}}
\put(1436.0,134.0){\rule[-0.200pt]{0.400pt}{4.818pt}}
\put(1436,89){\makebox(0,0){250}}
\put(1436.0,835.0){\rule[-0.200pt]{0.400pt}{4.818pt}}
\put(197.0,134.0){\rule[-0.200pt]{298.475pt}{0.400pt}}
\put(1436.0,134.0){\rule[-0.200pt]{0.400pt}{173.689pt}}
\put(197.0,855.0){\rule[-0.200pt]{298.475pt}{0.400pt}}
\put(45,494){\makebox(0,0){${v({\bf s})}\over{v_d}$}}
\put(816,44){\makebox(0,0){N}}
\put(197.0,134.0){\rule[-0.200pt]{0.400pt}{173.689pt}}
\put(212,144){\raisebox{-.8pt}{\makebox(0,0){$\Diamond$}}}
\put(217,194){\raisebox{-.8pt}{\makebox(0,0){$\Diamond$}}}
\put(222,239){\raisebox{-.8pt}{\makebox(0,0){$\Diamond$}}}
\put(227,291){\raisebox{-.8pt}{\makebox(0,0){$\Diamond$}}}
\put(232,338){\raisebox{-.8pt}{\makebox(0,0){$\Diamond$}}}
\put(445,667){\raisebox{-.8pt}{\makebox(0,0){$\Diamond$}}}
\put(693,690){\raisebox{-.8pt}{\makebox(0,0){$\Diamond$}}}
\put(940,702){\raisebox{-.8pt}{\makebox(0,0){$\Diamond$}}}
\put(1188,706){\raisebox{-.8pt}{\makebox(0,0){$\Diamond$}}}
\end{picture}
\caption{
\label{u_ten}
Strong field $NE=10$ data. $L/N$, $S^2/N$, and $v(\tube)/v_d$ are 
shown, respectively, in parts a, b and c.}
\end{figure}

Figures~\ref{u_one} and~\ref{u_ten} display our results for
$\overline{L}/(N-1)$ and $\overline{S^2}/(N-1)$, and the ratio
$v(\tube)/v_d$ calculated by inserting average values of $L$ and $S^2$
into equation~(\ref{weak_field_velocity}) for the velocity in weak
fields.  We divide $L$ and $S^2$ by N-1, because the finite size
correction then vanishes in the limit $u=NE~\rightarrow 0$.  Data for
small N=3-7 are exact matrix calculations, while data for large $N$
are simulated.

It appears that $\overline{L}/(N-1)\rightarrow 2/3$ for any value of
the scaling variable $u$. Dashed lines in figures~\ref{u_one}a
and~\ref{u_ten}a indicate the value $2/3$. The finite size correction
falls off proportionally to $u/N$. Our analysis of the
scaling limit in section (3) explains these results. Assuming that
$\overline{S^2}$ equals $N$ times a function of $u$,
equation~(\ref{weak_field_fugacity}) predicts the fugacity takes its
zero-field value with finite-size corrections of order $E=u/N$. The
stored length density $N_0/L$, which depends on the fugacity, inherits
the zero-field value and $u/N$ finite-size correction and in turn
passes them on to $\overline{L}/(N-1)$.

In contrast, the asymptotic value of $\overline{S^2}$ depends on the
scaling variable $u$. Variation of the end-to-end separation causes
the entire functional dependence of drift velocity on $E$ and $N$,
because $\overline{N_0}$ and $\overline{L}$ are constants in the
scaling limit.  References~\cite{BRF} and~\cite{BMW} predict a
cigar-shape polymer conformation for large values of $u$, with
elongation parallel to the field of order $N |E|^{1/2}$. Assuming that
the head and tail of the polymer lie near the top and bottom of this
cigar shape, we identify the elongation with our variable $S$.  From
equation (\ref{weak_field_velocity}) it follows that the large $u$
drift velocity varies as $E|E|$ independent of $N$, consistent with
the large $u$ form of $J(u)$.

Examining parts $c$ of figures~\ref{u_one} and~\ref{u_ten}, we
see that in the large $N$ limit, the velocity predicted by the weak
field limit~(\ref{weak_field_velocity}) converges to the actual
simulated velocity.  The large $N$ behavior is consistent with the
ratio of equation~(\ref{weak_field_velocity}) to simulated velocity
approaches $1$ as $1/N^{\sigma}$ with the power $\sigma$ falling
between $1/2$ and $1$.  This holds for small, moderate and large
values of the scaling variable, because finite-size and -field
corrections vanish in the scaling limit.

In conclusion, we present a general formula~(\ref{v_tube_general}) for
the drift velocity of any tube conformation in any electric field. The
simpler result~(\ref{weak_field_velocity}) applies throughout the
scaling limit of long chains at fixed $u=NE$. Our approach works
because diffusion of stored length decouples from fluctuations of
chain conformation except over a negligible length near the ends of
the chain.  The velocity is governed entirely by the end-to-end
separation of the tube S.  Tube fluctuations~\cite{BRF} are important
only because they permit dependence of $\overline{S^2}$ on $u$.

\newpage
\begin{ack}
	We acknowledge useful discussions with Gerard Barkema, John
Marko, Mark Newman and Ben Widom. We are especially indebted to Gerard
Barkema for sharing his simulation program with us. This work was
supported by NSF grant DMR-9221596 and by a Saudi Arabian government
fellowship.
\end{ack}

\end{document}